\documentclass{superfri} 

\usepackage{color}
\usepackage{datetime}
\usepackage{wrapfig}
\usepackage[framemethod=tikz]{mdframed}

\newcommand{\team}[1] {}
\newcommand{\pagebudget}[1] {}

\newcommand{\instruction}[1]{} 

\newcommand{\commentout}[1]{}

\begin{document}

\author{
Roscoe Bartlett\footnote{\label{inst1}Sandia National Laboratories} \and
Irina Demeshko\footnote{\label{inst2}Los Alamos National Laboratory} \and
Todd Gamblin\footnote{\label{inst3}Lawrence Livermore National Laboratory} \and
Glenn Hammond\footnoteref{inst1} \and
Michael Heroux\footnoteref{inst1} \and
Jeffrey Johnson\footnote{\label{inst4}Salesforce} \and
Alicia Klinvex\footnoteref{inst1} \and
Xiaoye Li\footnote{\label{inst6}Lawrence Berkeley National Laboratory} \and
Lois Curfman McInnes\footnote{\label{inst5}Argonne National Laboratory} \and
J.~David Moulton\footnoteref{inst2} \and
Daniel Osei-Kuffuor\footnoteref{inst3} \and
Jason Sarich\footnoteref{inst5} \and
Barry Smith\footnoteref{inst5} \and
Jim Willenbring\footnoteref{inst1} \and
Ulrike Meier Yang\footnoteref{inst3} 
}

\title{xSDK Foundations: Toward an Extreme-scale Scientific Software Development Kit}

\maketitle{}



\begin{abstract}
Extreme-scale computational science increasingly demands multiscale and multiphysics formulations. Combining software developed by independent groups is imperative: no single team has resources for all predictive science and decision support capabilities. Scientific libraries provide high-quality, reusable software components for constructing applications with improved robustness and portability.  However, without coordination, many libraries cannot be easily composed.  Namespace collisions, inconsistent arguments, lack of third-party software versioning, and additional difficulties make composition costly.

The Extreme-scale Scientific Software Development Kit (xSDK) defines community policies to improve code quality and compatibility across independently developed packages (hypre, PETSc, SuperLU, Trilinos, and Alquimia) and provides a foundation for addressing broader issues in software interoperability, performance portability, and sustainability.  The xSDK provides turnkey installation of member software and seamless combination of aggregate capabilities, and it marks first steps toward extreme-scale scientific software ecosystems from which future applications can be composed rapidly with assured quality and scalability.

\keywords{xSDK, Extreme-scale scientific software development kit, numerical libraries, software interoperability, sustainability}
\end{abstract}

\section{Software Challenges for Extreme-scale Science \pagebudget{1.0}}
\label{sec:intro}

Extreme-scale architectures provide unprecedented resources for
scientific discovery.  At the same time, the computational science and engineering (CSE) community faces daunting productivity and
sustainability challenges for parallel application development
\cite{
SoftwareProductivityWorkshopReport14,
CSESSP16,
RuedeWillcoxMcInnesDeSterckEtAl2016AllAuthors,
ASCR-ExascaleRequirementsReview16}.
Difficulties include increasing complexity of algorithms and computer
science techniques required by coupled multiscale and multiphysics
applications.  Further complications come from the imperative of portable performance in the midst of
dramatic and disruptive architectural changes on the path to exascale,
the realities of large legacy code bases, and human factors arising
in distributed multidisciplinary research teams pursuing leading edge
parallel performance.  Moreover, new architectures require fundamental algorithm
and software refactoring, while at the same time demand is
increasing for greater reproducibility of simulation and analysis
results for predictive science.  

This confluence of challenges brings with it a unique opportunity to fundamentally
change how scientific software is designed, developed, and
sustained. The demands arising from so many challenges force the CSE community to consider a broader range of potential solutions.  It is this setting that makes possible a collaborative effort to establish a scientific software ecosystem of reusable libraries and community policies to guide common adoption of practices, tools, and infrastructure. Incremental change is not a viable option, so migration to a new model for CSE software is possible.

The xSDK has emerged as a first step toward a new ecosystem, where application codes are composed via interfaces from a common base of reusable components more than they are developed from a clean slate or derived from monolithic code bases.  To the extent that this compositional approach can be reliably used, new CSE applications can be created more rapidly, with greater robustness and scalability, by smaller teams of scientists, enabling them to focus more attention on obtaining science results than on the incendentals of their computing environment.

\subsection{Related work \pagebudget{1.0}}


The scientific software community has a rich tradition of defining \textit{de facto} standards for collections of capabilities.  EISPACK~\cite{EISPACK,EISPACKEXT}, LINPACK~\cite{LINPACK}, 
BLAS~\cite{BLAS1,BLAS1a,BLAS2,BLAS3}, and LAPACK~\cite{LAPACK} delivered a sound foundation for numerical linear algebra in libraries and applications.  Commercial entities such as the Numerical Algorithms Group (NAG)~\cite{nag:homepage}, the Harwell Subroutine Library (HSL)~\cite{hsl:homepage} and IMSL~\cite{imsl:homepage} have provided high quality, unified software capabilities to users for decades. 

More recently, the TOPS~\cite{tops:project}, 
ITAPS~\cite{itaps:project}, and FASTMath~\cite{fastmath:project} SciDAC institutes brought together developers of large-scale scientific software libraries. While these libraries were independently developed by distinct teams and version support lacked coordination,
the collaborations sparked exchange of experiences and discussion of practices that avoided potential pitfalls and facilitated the combined use of the libraries\cite{MillerEtAl2013} as needed by scientific teams.
Prior efforts to provide interoperability between solver libraries can be found in PETSc \cite{petsc:homepage}, which allows users to access libraries such as hypre~\cite{hypre:homepage} and SuperLU~\cite{superlu:homepage} by using the PETSc interface, sparing users the effort to rebuild their problems through hypre's or SuperLU's interfaces. Trilinos~\cite{trilinos:homepage}, a collection of self-contained software packages, also provides ways for users to gain uniform access to third-party scientific libraries.  


\section{xSDK Vision \pagebudget{2.0}}
\label{sec:vision}

The complexity of application codes is steadily increasing due to more sophisticated scientific models and the continuous emergence of new high-performance computers, making it crucial to develop software libraries that provide needed capabilities and continue to adapt to new computer architectures. Each library is complex and requires different expertise.  Without coordination, and in service of distinct user communities, this circumstance has led to difficulties when building application codes that use 8 or 10 different libraries, which in turn might require additional libraries or even different versions of the same libraries. 

The xSDK represents a different approach to coordinating library development and deployment.  Prior to the xSDK, scientific software packages were cohesive with a single team effort, but not across these efforts. The xSDK goes a step further by developing community policies followed by each independent library included in the xSDK.  This policy-driven, coordinated approach enables independent development that still results in compatible and composable capabilities.

\begin{figure}[h!]
	\begin{center}
		\includegraphics[width=0.99\linewidth]{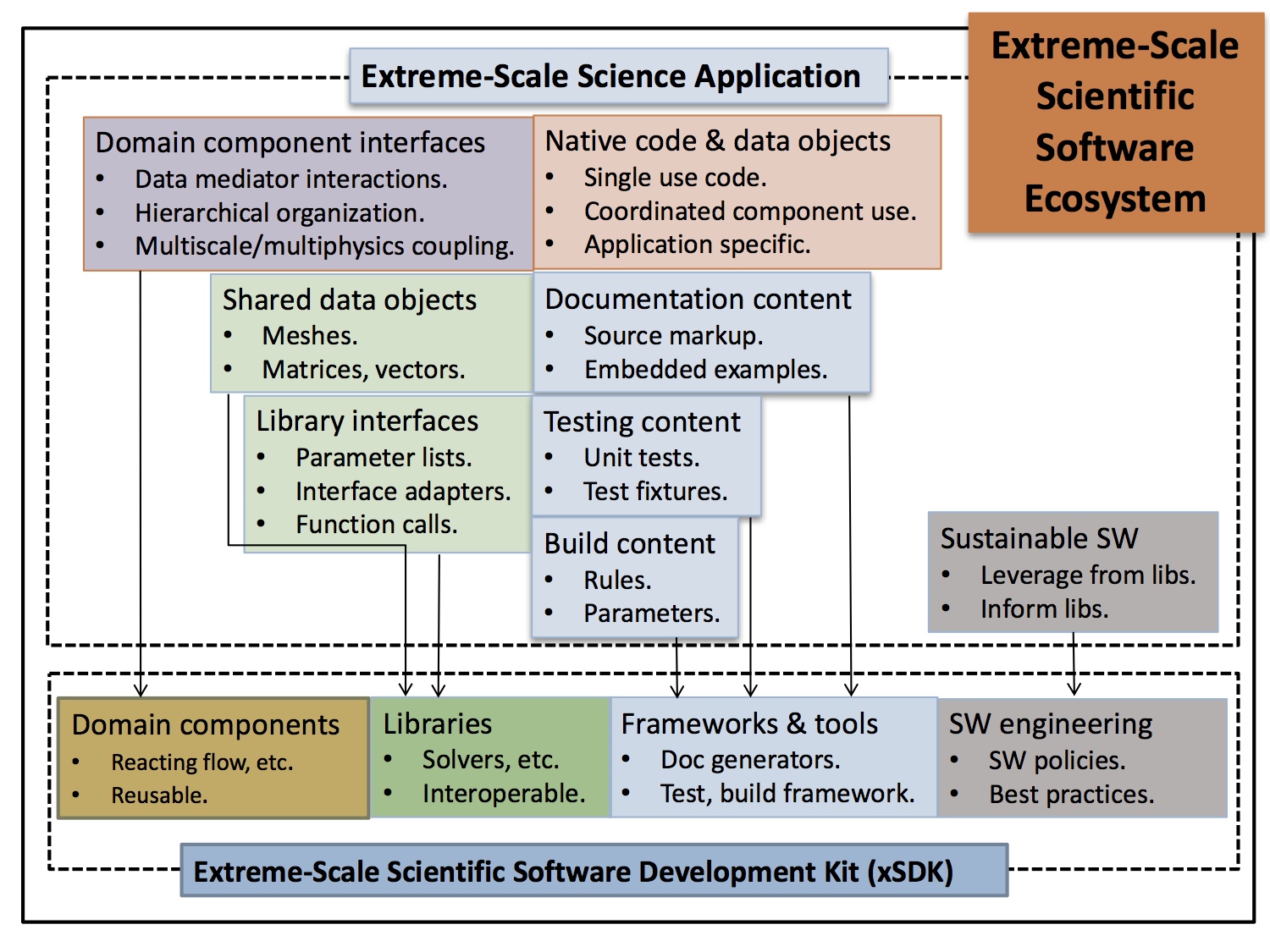}  \hspace{1pc}
		\caption{ \label{fig:BigPicture}\itshape\small
The xSDK intends to provide the foundation for a modern extreme-scale scientific software
ecosystem, where application development is accomplished by composition of
high-quality, reusable software components rather than by tangential use of
libraries.  Application developers produce a small portion of custom code that
expresses the particular purpose of the software and then gain the bulk of
functionality by parameterized use of xSDK components and libraries, which are
developed by diverse, independent groups throughout the community.  xSDK
frameworks for documentation, testing, and code quality, as well as established
software policies and best practices, can be adapted and adopted as appropriate
by the application developers to provide compatible, high-quality, and
sustainable software. As we move toward this new ecosystem, application
development times from first concept to scalable production code should drop
dramatically.  Success hinges on the quality, interoperability, usability, and
diversity of xSDK capabilities and our ability to deliver the xSDK to domain
scientists.
}
	\end{center}
\end{figure}

The initial xSDK project is the first step toward a comprehensive software ecosystem.
As shown in Figure~\ref{fig:BigPicture}, the vision of the xSDK is to
provide infrastructure for and interoperability of a collection of
related and complementary software elements---developed by diverse,
independent teams throughout the high-performance computing (HPC)
community---that provide the building blocks, tools, models,
processes, and related artifacts for rapid and efficient development
of high-quality applications.  Our long-term goal is to make the xSDK
a turnkey standard software ecosystem that is easily installed on
common computing platforms, and can be assumed as available on any
leadership computing system in the same way that BLAS and LAPACK are
available today.

\subsection{Elements of an Extreme-scale Scientific Software Ecosystem}

Rapid, efficient production of high-quality, sustainable applications
is best accomplished using a rich collection of reusable libraries,
tools, lightweight frameworks, and defined software methodologies,
developed by a community of scientists who are striving to identify,
adapt, and adopt best practices in software engineering.  Although the
software engineering community has ongoing debate about the precise
meaning of terms, we define the basic elements of a scientific software 
ecosystem to include:
\begin{itemize}
\item \textbf{Library:} High-quality, encapsulated, documented, tested and
multi-use software that is incorporated into the application and used as native
source functionality.  Libraries can provide \textit{control inversion} via
abstract interfaces, call-backs, or similar techniques such that user-defined
functionality can be invoked by the library, e.g., a user-defined sparse matrix
multiplication routine. Libraries can also provide factories that facilitate
construction of specific objects that are related by a base type and later used
as an instance of the base type. Libraries can include domain-specific software
components that are designed to be used by more than one application.

\item \textbf{Domain component:} Reusable software that is intended for modest
reuse across applications in the same domain.  Although this kind of component
is a library, the artifacts and processes needed to support a component are
somewhat different than for a broadly reusable library.

\item \textbf{Framework:} A software environment that implements specific design
patterns and permits the user to insert custom content.  Frameworks include
documentation, build (compilation), and testing environments. These frameworks are
lightweight and general purpose.  Other frameworks, such as multiphysics, are 
considered separately, built on top of what we describe here.

\item \textbf{Tool:} Software that exists outside of applications,
  used to improve quality, efficiency, and cost of developing and
  maintaining applications and libraries.

\item \textbf{Software development kit (SDK):} A collection of related
  and complementary software elements that provide the building
  blocks, libraries, tools, models, processes, and related artifacts for rapid and
  efficient development of high-quality applications.

\end{itemize}
\vspace{.1in}
Given these basic elements, we define an application code as the following composition:
\begin{itemize}

\item \textbf{Native data and code:} Every application will have a
  primary routine (often a main program) and its own collection of source code and private data.
  Historically, applications have been primarily composed of native
  source and data, using libraries for a small portion of
  functionality, such as solvers.  We foresee a decrease in the amount
  of native code required to develop an application by extracting and
  transforming useful native code into libraries and domain components, making it available  to other applications.


\item \textbf{Component and library function calls:} Some application
functionality is provided by invoking library functions.  We expect to increase
usage of libraries as a part of our efforts.

\item \textbf{Library interface adapters:} Advanced library integration often
involves invoking the control inversion facilities of the library in order to
incorporate application-specific knowledge.  In the case of sensitivity
analysis, embedded optimization, and related analyses, control inversion via
these adapters is essential in order to permit the solver to invoke the
application with specific input data.

\item \textbf{Component and library parameter lists:} Libraries tend to provide
a broad collection of functionality for which parameters must be set.

\item \textbf{Shared component and library data:} Most libraries require the
user to provide nontrivial data objects, such as meshes or sparse matrices, and
may provide functions to assist the application in constructing these objects. 
Unlike parameter list definitions, which represent a narrow interface dependency
between the application and library, application-library data interfaces can be
very complicated.

\item \textbf{Documentation, build, and testing content:} The
  application-specific text, data, and source used by the
  documentation, build, and testing frameworks to produce the derived
  software documentation, compilation, and test artifacts.

\end{itemize}

\section{xSDK Approach \pagebudget{0.5}}
\label{sec:approach}

The xSDK approach to developing software has two distinguishing features from previous efforts in the scientific computing community:
\begin{itemize}
	\item \textbf{Peer-to-peer interoperability:} Some previous efforts\footnote{A notable example is the Equation Solver Interface (ESI), which defined an abstraction layer to present a common client interface to distinct software products. The challenge of this approach is that the unique features of the underlying products were difficult to access.  The very use of a common abstraction reduced the usability of these products.} attempted to use additional abstraction layers that would hide differences in the underlying packages.  The xSDK approach uses the existing extensibility features of the libraries to enable peer-to-peer access of capabilities at various levels of interoperability through the native interfaces of the packages.
For example, if a user has already integrated PETSc data structures into their code, the xSDK approach preserves that approach, but permits use of capabilities in hypre, SuperLU, and Trilinos with PETSc.
		
    \item\textbf{Software policies:} Most existing scientific software efforts rely on close collaboration of a single team in order to assure that collective efforts are compatible and complementary.  The xSDK relies instead on \textit{policies} that promote compatibility and complementarity of independently developed software packages.  By specifying only certain expectations for how software is designed, implemented, documented, supported, and installed, the xSDK enables independent development of separate packages, while still ensuring complementarity and composability.
\end{itemize}

The xSDK can assure interoperability and compliance with community policies because the leaders and developers of xSDK packages are members of the xSDK community.  If interface changes are required in a package or a version of a third-party solver needs to be updated, these changes will be made in the member package.  For example, in order for Trilinos and PETSc to use the same version of SuperLU and hypre, the Trilinos and PETSc developers commit to agreeing on changes to Trilinos and PETSc that are needed for compatibility.  Similarly, changes to interfaces for interoperability and inversion of control (see the next  Section~\ref{sec:interoperability}) are done within the xSDK packages, and regularly tested for regressions.  xSDK interoperability is possible because of the commitment of xSDK member package development teams.



\subsection{xSDK library interoperability}
\label{sec:interoperability}

A fundamental objective of the xSDK project is to 
provide interoperability layers among hypre,
PETSc, SuperLU, and Trilinos packages, as appropriate,
with the ultimate goal of making all mathematically meaningful
interoperabilities possible in order to fully support exascale applications.

Software library interoperability refers to the ability of two or more libraries
to be used together in an application, without special effort by the
user~\cite{InteroperableLibraries:xSDK:2016}.
For simplicity, we discuss interoperability between two libraries; extension 
to three or more libraries is conceptually straightforward.
Depending on application needs, various levels of interoperability can be considered:
\begin{itemize}
\item {\em Interoperability level 1}: both libraries can be used (side by side)
in an application
\item {\em Interoperability level 2}: both libraries can exchange data (or
control data) with each other
\item {\em Interoperability level 3}: each library can call the other library to
perform unique computations
\end{itemize}

The simplest case (interoperability level 1) occurs when an
application needs to call two distinct libraries for different
functionalities (for example, an MPI library for message-passing
communication and HDF5 for data output). As discussed in
\cite{MillerEtAl2004,MillerEtAl2013}, even this basic 
interoperability requires consistency among
libraries to be used in the same application, in terms of
compiler, compiler version/options, and third-party
capabilities. 
If both libraries have a dependency on a common third
party, the libraries must be able to use a single common instance of
it. For example, more than one version of the popular SuperLU linear
solver library exists, and interfaces have evolved. If two libraries
both use SuperLU, they must be able to work with the same version of
SuperLU. 
In practice, installing multiple 
independently developed
packages together can be a tedious trial-and-error process.  
The definition and implementation of 
xSDK community policies
standards have overcome this difficulty for xSDK-compatible packages.

Interoperability level 2 builds on level 1 by enabling conversion, or
encapsulation, and exchange of data between libraries. This level can
simplify use of libraries in sequence by an application.  In this
case, the libraries themselves are typically used without internal
modification to support the interoperability.  
Future work on node-level resource management is essential 
to support this deeper level of software interoperability
for emerging architectures.

Interoperability level 3 builds on level 2 by supporting the use of
one library to provide functionality on behalf of another
library. This {\em integrated execution} provides significant value to
application developers because they can access capabilities of
additional libraries through the familiar interfaces of the first
library.

The remainder of this section discusses proposed work on integrated execution, where
our guiding principles are to provide interoperability that is
intuitive and easy to use, and to expose functionality of each library where feasible.

\subparagraph{Control inversion.}

Interoperability between two (or more) existing library components can
be achieved by one of two basic mechanisms: (i) create an abstraction
layer that sits on top of both components to act as an intermediary
between the user and both components or (ii) permit users to write
directly to the interface of one component and provide peer-level
interoperability between the two components. For example, consider the
matrix construction capabilities in PETSc and Trilinos. Both libraries
provide extensive support for piecewise construction of sparse
matrices, as needed for building objects in applications based on finite
elements/volumes/differences. It would be possible, in
principle, to create a top-level abstraction layer that could be used
to build a sparse matrix or other data objects for PETSc or Trilinos,
depending on an input option to select either target. Alternatively,
the user can construct the data object by using the PETSc or Trilinos
functions directly, and then we can create adapters in Trilinos and
PETSc to wrap the respective matrix object and make it behave like one
of its own.

Although the first approach may seem attractive, it is difficult to
develop in a sustainable and effective way. PETSc and Trilinos data
object construction processes are targeted to specific programming,
language, and usage models. The differences in approach may appear
small, but are very important in terms of developer productivity, code
portability, and expressiveness. Any abstraction layer that would sit
on top of both would discard the simplicity of one approach or the
expressiveness of the other.

Peer-to-peer interoperability is much more attractive than a general
abstraction layer. The  xSDK libraries have mechanisms to work with or
easily transform existing data objects that were built outside their
own construction processes. For example, a PETSc sparse matrix can be
used within Trilinos, without copying, by using an adapter class. A
similar approach can work with a Trilinos matrix used by PETSc.

The hypre and SuperLU libraries do not directly support control inversion in the same way as PETSc and Trilinos, but do advertise their input data structures such that PETSc and Trilinos can construct compatible data structures that are passed to hypre and SuperLU without copying.

The current release of xSDK does not support all possible opportunities for interoperability.  Level 1 interoperability is complete within the current xSDK.  Level 2 interoperability is partial, with Trilinos being able to accept PETSc data structures.  Level 3 interoperability is also partially available with PETSc and Trilinos able to call use hypre and SuperLU.


\subsection{xSDK community policies \pagebudget{2.0}}
\label{sec:policies}

In \cite{MillerEtAl2013,MillerEtAl2004} various software quality engineering practices for `smart libraries' are discussed that, when followed, can alleviate generation of an application executable that depends on many libraries, reduce mistakes in how to use these libraries, and provide help to users to identify and correct errors when they occur.

The first xSDK release
demonstrated the impact
of defining xSDK commmunity policies, including standard GNU
autoconf and CMake options to
simplify the combined use, portability, and sustainability of
independently developed software packages (hypre, PETSc, SuperLU,
and Trilinos) and provide a foundation for addressing broader issues
in software interoperability and performance portability.
\begin{figure}[h!]
 \vspace{1pc}
 \begin{mdframed}
	\begin{center}
		{\bf xSDK Mandatory Policies}
	\end{center}
\textbf{Must:}
\begin{itemize}
		\item[M1.] Support xSDK community GNU Autoconf or CMake options~\cite{xSDK-community-installation-policies2016}.
		\item[M2.] Provide a comprehensive test suite.
		\item[M3.] Employ user-provided MPI communicator.
		\item[M4.] Give “best effort” at portability to key architectures.
		\item[M5.] Provide a documented, reliable way to contact the development team.
		\item[M6.] Respect system resources and settings made by other previously called packages.
		\item[M7.] Come with an open source license.
		\item[M8.] Provide a runtime API to return the current version number of the software.
		\item[M9.] Use a limited and well-defined symbol, macro, library, and include file name space.
		\item[M10.] Provide an accessible repository (not necessarily publicly available).
		\item[M11.] Have no hardwired print or IO statements.
		\item[M12.] Allow installing, building, and linking against an outside copy of external software.  
		\item[M13.] Install headers and libraries under \verb|<prefix>/include/| and \verb|<prefix>/lib/|.
		\item[M14.] Be buildable using 64 bit pointers. 32 bit is optional.
\end{itemize}
	\begin{center}		
		{\bf xSDK Recommended Policies}
	\end{center}
\textbf{Should:}
\begin{itemize}	
		\item[R1.] Have a public repository.
		\item[R2.] Possible to run test suite under valgrind in order to test for memory corruption issues.
		\item[R3.] Adopt and document consistent system for error conditions/exceptions.
		\item[R4.] Free all system resources it has acquired as soon as they are no longer needed.
		\item[R5.] Provide a mechanism to export ordered list of library dependencies.
\end{itemize}
\end{mdframed}
 \hspace{1pc}
		\caption{ \label{fig:CommunityPoliciesBrief}\itshape\small
xSDK community policies specify expectations that any software library or framework (henceforth referred to as package) must satisfy in order to be xSDK compatible.  The designation of a package being xSDK compatible informs potential users that the package can be easily used with other xSDK libraries and components and thus helps to address issues in long-term sustainability and interoperability among packages.			
}

\end{figure}

xSDK community package policies~\cite{xSDK-community-package-policies2016},
briefly summarized in Figure~\ref{fig:CommunityPoliciesBrief},
are a set of minimum requirements (including topics of configuring, installing,
testing, MPI usage, portability, contact and version information, open source
licensing, namespacing, and repository access) that a software package must
satisfy in order to be considered xSDK compatible. The designation of xSDK
compatibility informs potential users that a package can be easily used with others.

xSDK community installation policies~\cite{xSDK-community-installation-policies2016}
help make configuration and installation of xSDK software and other HPC packages
as efficient as possible on common platforms, including standard Linux
distributions and Mac OS X, as well as on target machines currently available at
DOE computing facilities (ALCF, NERSC, and OLCF) and eventually on new exascale platforms.

Community policies for the xSDK promote long-term sustainability and
interoperability among packages, as a foundation for supporting
complex multiphysics and multiscale ECP applications. In addition,
because new xSDK packages will follow the same standard, installation
software and package managers (for example, Spack~\cite{gamblin+:sc15}) can easily be
extended to install many packages automatically.

\begin{wrapfigure}{R}{0.70\textwidth}
\vspace{-0.3in}
\includegraphics[width=0.70\textwidth]{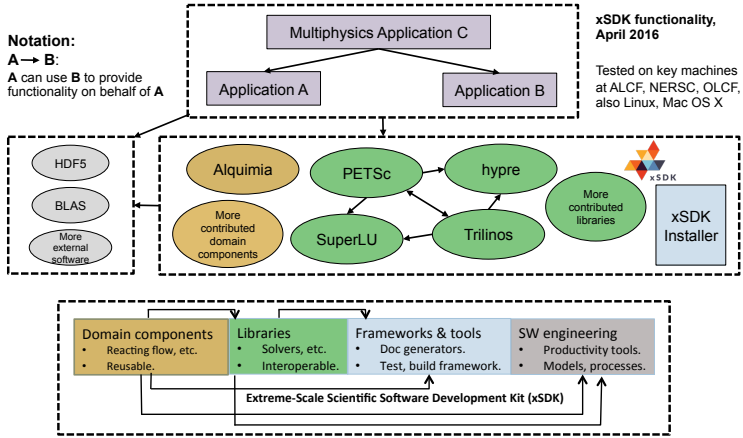} 
\vspace{-1.1in}
\caption{\textit{xSDK schematic diagram.}}
\label{fig:xsdk-schematic}
\end{wrapfigure}

Figure~\ref{fig:xsdk-schematic} illustrates a new {\em Multiphysics
Application C}, built from two complementary applications that can
readily employ any libraries in the xSDK (hypre, PETSc, SuperLU, and
Trilinos, shown in green).  Application domain components are represented
in orange.  Of particular note is Alquimia~\cite{alquimia:homepage}, a domain-specific interface 
that support uniform access to multiple biogeochemistry capabilities, including
PFLOTRAN~\cite{pflotran:homepage}.
The arrows among
the xSDK libraries indicate current support for
a package to call another to provide scalable linear solvers
functionality on its behalf.  For example, {\em Application~A} could
use PETSc for an implicit-explicit time advance, which in turn could
interface to SuperLU to solve the resulting linear systems with a
sparse direct solver.  {\em Application~B} could use Trilinos to solve
a nonlinear system, which in turn could interface to hypre to solve
the resulting linear systems with algebraic multigrid.  Of course,
many other combinations of solver interoperability are also possible.  
The website \url{https://xsdk.info/example-usage} 
and \cite{Klinvex-xSDKTrilinos} provide examples of
xSDK usage, including interoperability among linear solvers in hypre,
PETSc, SuperLU, and Trilinos.

\subsection{xSDK coordinated software releases and current status \pagebudget{1.0}}
\label{sec:releases}

%

The first critical step in producing the initial release of the xSDK in April 2016 was to define what the release should look like. On the loosely coupled end of the spectrum, one possibility would have been to certify that specific release versions of the different packages are compatible with one another and not coordinate the distribution of the release beyond that. On the other end of the spectrum were possibilities such as a common test or even build infrastructure. The xSDK team decided on a strategy that provided a single point of distribution for all xSDK component packages, but did not force a common infrastructure on the packages, beyond the agreed upon community policies.

The chosen distribution mechanism was, at its core, the existing PETSc distribution mechanism~\cite{petsc:homepage}. This was a stable, well-supported option that required a relatively small amount of effort to extend for the xSDK use case, because it already supported the majority of xSDK component packages, and very little additional ongoing maintenance beyond what the PETSc team was already doing. The latter was a key consideration because, when possible, the xSDK team actively avoids solutions that create long-term maintenance beyond what is needed for the component packages.

In addition to creating and updating interfaces between xSDK packages to provide new functionality, significant work was done to make it possible to install all of the xSDK packages together. For example, PETSc and Trilinos depended on different versions of SuperLU, and no single version of SuperLU could be used with both the PETSc and Trilinos SuperLU interfaces enabled.

As the coordinated release effort began (initially work began with package-to-package compatibility and interface efforts), the xSDK installer was used to test release versions of the component packages together to avoid the churn of the various development versions. Testing was set up by multiple xSDK component package teams using their own resources. Again, sustainability was a focus. Having the individual teams manage separate test builds in which each team had a vested interest was a better choice than a centralized effort that would have no clear owner in the absence of an xSDK-level funding source. The test builds included release and development versions, but leading up to the release, primary focus was given to the release testing.

The existing release processes of the various xSDK component packages varied greatly in terms of testing and overall rigor. However, since each component package had a release process that the individual development teams had determined was sufficient for their needs, the decision was made to focus on requirements outside of typical release requirements. Specifically, this involved providing the name of a branch to use for release candidate testing, setting up tests for xSDK-level build configurations most relevant to the component package, and being responsive to any issues found.

In addition to the testing conducted by each package team, the xSDK 0.1 release was ported to three target platforms at three different computing facilities: Mira at ALCF, Edison at NERSC, and Titan at OLCF. One developer was primarily responsible for each of the three porting efforts, and those people coordinated with other xSDK developers and component package team developers 
to resolve porting issues as necessary.

The official tag for the initial xSDK release was chosen to be v0.1.0. After the initial release, a patch release, v0.1.1, was completed. The versions of the component packages used for this subsequent release were either the same version used for the initial 0.1 release, or a patch release of the component package based on the release used for the initial release. According to xSDK release policy, only patch-level updates for component packages are to be used for xSDK patch releases, and only patch or minor release version updates for component packages are to be used for xSDK minor releases.

Prior to selecting the PETSc distribution capability for the xSDK 0.1 release, Spack~\cite{gamblin+:sc15} was also seriously considered. Spack is a multi-platform package manager that supports a variety of compilers, libraries, and applications, as well as the installation of multiple concurrent versions and software configurations. Because of the increasing popularity and robustness of Spack, and the need to expand the xSDK to include several additional component packages, the xSDK team decided to use Spack to build an alpha release version of the xSDK, to be released in early 2017. Going forward, the intent is to use Spack as the principal supported xSDK distribution capability.

\section{Next Steps}
The first xSDK releases focused on discovery of collaboration models and community building among four of the major open source scientific library projects in the international scientific computing community: hypre, PETSc, SuperLU and Trilinos.  The xSDK project will continue over the next few years under the United States Department of Energy Exascale Computing Project (ECP)~\cite{ecp:homepage}.

Our efforts so far have established a baseline for expanding the xSDK scope under ECP funding in several important directions:
\begin{enumerate}
	\item \textbf{Include more libraries:} The xSDK will expand to include all library efforts under ECP funding.  Spcifically, widely used libraries such as SUNDIALS and Magma will become xSDK compatible, as will new efforts that address the performance challenges of exascale computing platforms.

	\item \textbf{Further refine and expand community policies:} While the current xSDK community policies, summarized in Figure~\ref{fig:CommunityPoliciesBrief}, are extremely useful as a mechanism to improve interoperability and compatibility of independently developed scientific libraries, we believe we can further refine and expand these policies
to better assure software quality and further realize the scientific software ecosystem sketched in Figure~\ref{fig:BigPicture}.

	\item \textbf{Include more domain components:} As described in Section~\ref{sec:intro}, the vision of the xSDK is to create a software ecosystem where new scientific applications are composed via interfaces from a common base of reusable domain components and libraries.  We will work with science teams to identify opportunities for creating collections of domain components for their communities.
	
	\item \textbf{Explore the use of community installation tools, including Spack:} While the extended PETSc installer has been very useful for establishing xSDK as a unified project, Spack~\cite{gamblin+:sc15} promises to provide a tool that serves and is supported by a larger community, making it very appealing as the principal long term installation tool for xSDK libraries.
	
	\item \textbf{Process control transfer interfaces:} The ever-increasing use of concurrency within the top-level MPI processes requires that computational resources used by an application or library can be transferred to another library. Transfer of these resources is essential for obtaining good performance.  The xSDK project will develop interfaces to support sharing and transfer of computational resources.
	
\end{enumerate}

\section{Conclusions \pagebudget{1.0}}
\label{sec:conclusions}

The extreme-scale scientific community faces numerous disruptive challenges in the coming decade.  Fundamental limits of physics are forcing changes that dramatically impact all system layers from architecture to application software design.  These disruptive changes drive us to move beyond incremental change in scientific application design and implementation.  Establishing a scientific software ecosystem that focuses more on the composition of scalable, reusable components for application software development can provide an attractive alternative and the xSDK is the first step toward the ecosystem described in Figure~\ref{fig:BigPicture}.

Community policies for the xSDK promote long-term sustainability and interoperability among packages, as a foundation for supporting complex multiscale and multiphysics applications. The designation of xSDK compatibility informs potential users that a package can be easily used with other xSDK libraries and components. In addition, because new xSDK packages will follow the same standard, installxSDK and package managers can easily be extended to install many packages automatically.

Interoperability of xSDK member packages, when wrapped with adequate testing, enables a sustainable coupling of capabilities that enable applications to use xSDK member packages as a cohesive suite. The first xSDK releases demonstrate the impact of xSDK community policies, testing and examples to simplify the combined use, interoperability, and portability of independently developed software packages, establishing the first step toward realizing an extreme-scale scientific software ecosystem.

\ack{This material is based upon work funded by the U.S. Department of Energy Office of Science, Advanced Scientific Computing Research and Biological and Environmental Research programs.  We thank program managers Thomas Ndousse-Fetter, Paul Bayer, and David Lesmes for their support.

The work of ANL authors is supported by the U.S. Department of Energy, Office of Science, under Contract DE-AC02-06CH11357. 
The work of LBNL authors is partially supported by the Director, Office of Science, Office of Advanced Scientific Computing Research of the US Department of Energy under contract no. DE-AC02-05CH11231.
Prepared by LLNL under Contract DE-AC52-07NA27344. 
Work of LANL authors is funded by the Department of Energy at Los Alamos National Laboratory under contract DE-AC52-06NA25396.
Sandia is a multiprogram laboratory operated by Sandia Corporation, a Lockheed Martin Company, for the United States Department of Energy's National Nuclear Security Administration under contract DEAC04-94AL85000.
}
\openaccess

\bibliography{bib/cited,bib/ideas}


\end{document}